\documentclass[twocolumn,prl,aps,amsmath,amssymb,floatfix,showpacs]{revtex4}
\usepackage{graphicx}
\usepackage{bm}

\begin{document}
\title{Electronic structure of heavily-doped graphene: the role of foreign atom states}

\author{Matteo Calandra}
\author{Francesco Mauri}
\affiliation{CNRS and 
Institut de Min\'eralogie et de Physique des Milieux condens\'es, 
case 115, 4 place Jussieu, 75252, Paris cedex 05, France}
\date{\today}

\begin{abstract}
Using density functional theory calculations we investigate the electronic structure
of graphene doped by deposition of foreign atoms. 
We demonstrate that, as the charge transfer
to the graphene layer increases, the band structure of the pristine graphene sheet is 
substantially affected. This is particularly relevant when Ca atoms are deposed
on graphene at CaC$_{6}$ stoichiometry.
Similarly to what happens in superconducting graphite
intercalated compounds, a Ca bands occurs at the Fermi level. Its
hybridization with the C states generates a strong non-linearity in one of the 
$\pi^{*}$ bands below the Fermi level, 
at energies comparable to the graphene E$_{2g}$ phonon
frequency. 
This strong non-linearity, and not manybody effects as
previously proposed,
explains the large and anisotropic values of the apparent 
electron-phonon coupling 
measured in angular resolved photoemission. 
\end{abstract}
\maketitle

The discovery of superconductivity in CaC$_{6} $\cite{Weller,Genevieve,Gauzzi} 
demonstrates that superconducting critical 
temperatures (T$_c$s) as large as 15 K can be 
obtained in Graphite Intercalated Compounds (GICs). Recently, 
deposition of foreign atoms onto a graphene monolayer \cite{Berger}
has been achieved. This finding  could
considerably widen the number of GICs,
since the constraints for deposition onto graphene
are milder than those required for graphite intercalation.
Consequently even larger T$_c$s could be discovered in these systems.  

So far, the electronic structure of atoms deposed on graphene has been interpreted
in terms of the pristine graphene band structure
\cite{Novoselov2005, ZhouPRB2005, Zhou2006NatPhys, Bostwick2007}.
This is correct if the charge transfer to the graphene layer
(doping) is weak.  However, this is questionable for larger dopings,
since foreign atom states could affect the band structure below the Fermi
level ($\epsilon_f$). Indeed in GICs, an 
intercalant band crosses $\epsilon_f$ \cite{Csanyi,Calandra2005,Calandra_GICs}.
As far as K deposition on graphene is concerned, 
detailed Angular Resolved Photo Emission (ARPES) measurements
\cite{Novoselov2005, ZhouPRB2005, Zhou2006NatPhys, Bostwick2007}
have shown that (i) the graphene band-structure below $\epsilon_f$
is weakly affected by the presence of K atoms,
(ii) a marked kink occurs at 0.195 eV below $\epsilon_f$, an energy
corresponding to the E$_{2g}$ graphene phonon frequency
and (iii) the electron-phonon coupling extracted from ARPES measurements
is 5.5 times larger than what could be expected on the basis
of a rigid doping of the graphene bands \cite{Calandra2005,Calandra2007}.
In a very recent paper \cite{McChesney2007} deposition of
Ca atoms on graphene at doping as large as that of CaC$_{6}$
was achieved and, most surprising, a massive enhancement of 
the electron-phonon coupling was reported.
The apparent electron-phonon coupling has been shown to be very anisotropic
with values ranging from 0.5 to 2.3, suggesting that an heavily-doped
graphene monolayer could be superconducting with large critical temperatures
and large anisotropic superconducting gap(s).
Such large and anisotropic values of the electron-phonon
coupling were interpreted as due to many-body effects and to the occurrence
of a Van hove singularity. However neither measurements nor
calculations of the electronic density of states were performed.

In this work we use density functional theory calculations to
interpret the huge and highly anisotropic electron-phonon
coupling observed in heavily doped graphene by ARPES in terms of the
CaC$_6$ monolayer band structure.

In ARPES the spectral weight is measured,  namely, for a given band index,
\begin{eqnarray}
A({\bf k},\epsilon)=\frac{-2[\Sigma_{\rm all}^{''}({\bf k},\epsilon)]}
{\left[\epsilon - \epsilon_{{\bf k}} 
-\Sigma_{\rm all}^{'}({\bf k},\epsilon)\right]^2
+ \left[\Sigma_{\rm all}^{''}({\bf k},\epsilon)+\right]^2}
\label{eq:SW}
\end{eqnarray}
where $\Sigma_{\rm all}^{'}({\bf k},\epsilon)$ and 
$\Sigma_{\rm all}^{''}({\bf k},\epsilon)$
are the real and imaginary parts of the electron self-energy
$\Sigma_{\rm all}({\bf k},\epsilon)$, and $\epsilon_{\bf k}$ are the
single particle bands. The total 
electron self-energy includes contributions from 
all the interactions in the system.
The electron-phonon coupling parameter is:
\begin{eqnarray}
\lambda_{\bf k}=\left.\frac{\partial 
\Sigma^{'}({\bf k},\epsilon)}{\partial \epsilon}\right|_{\epsilon=\epsilon_f}
\label{eq:mass_enhancement}
\end{eqnarray}
where $\Sigma({\bf k},\epsilon)$ is the electron-phonon contribution  to
the electron self-energy and $\Sigma^{'}({\bf k},\epsilon)$ its real part. 
In what follows we assume that the dominant 
contribution to the electron self-energy is given by the electron-phonon coupling,
$\Sigma_{\rm all}({\bf k},\epsilon)\approx\Sigma({\bf k},\epsilon)$.

The maximum position in the spectral weight is given by the
relation 
$\epsilon^{\rm max}_{\bf k}-\epsilon_{\bf k}-
\Sigma^{'}({\bf k},\epsilon^{\rm max}_{\bf k})=0$.
Linearizing $\Sigma^{'}({\bf k},\epsilon)\approx-\lambda_{\bf k}\epsilon$ 
leads to
$\lambda_{\bf k}=\frac{\epsilon_{\bf k}}{\epsilon^{\rm max}_{\bf k}}-1$. 
{\it If the bare bands behave linearly} ,namely  
$\epsilon_k=\hbar v_{0} k$,
{\it and if the renormalized can be linearized 
at} $\epsilon_f$, 
$\epsilon^{\rm max}_{\bf k}=\hbar v_{\rm f} k$, 
\cite{Grimvall,Calandra2007} then $\lambda_{\bf k}$ 
and
\begin{equation}
\lambda_{\bf k}=\frac{v_{0}}{v_{f}}-1
\label{eq:lambda_eff}
\end{equation}
where, $v_f$ is obtained from a linear fit to the maximum
position in momentum distribution curves (MDCs) at energies close to 
$\epsilon_f$, while $v_{0}$ is obtained from a linear fit 
in an appropriate energy window below the kink where the bare
bands are linear. Note that if the bare bands are not linear then
Eq. \ref{eq:lambda_eff} is incorrect.

In experiments a finite resolution affects
substantially the energy window close to $\epsilon_f$ so that
the determination of $v_f$ is non-trivial \cite{Bostwick_Review}. The
the following quantity is then computed:
\begin{equation}
\lambda_{\rm ARPES}=\frac{v_0}{v_{\rm ARPES}}-1 \label{eq:lambda_arpes}
\end{equation}
where now $v_{\rm ARPES}$ is obtained from a linear fit to the maximum position
in MDCs in an energy window ranging from the kink energy up to the Fermi level.
Clearly $\lambda_{\rm ARPES}\approx\lambda$ only if $v_{f}\approx v_{\rm ARPES}$. 
Consequently the energy window chosen to fit $ v_{\rm ARPES}$ is crucial. This has
been demonstrated to be important in graphene at low electron-doping 
\cite{Calandra2007}. Indeed, in the case of rigid band doping of the graphene
$\pi^{*}$ bands, where the electron-phonon coupling can be calculated
analytically \cite{Calandra2005,Calandra2007},  it has been found that 
$\lambda_{\rm ARPES}\approx 2.5 \lambda$. Thus, in this case, $\lambda_{\rm ARPES}$
cannot provide a quantitative information of the electron-phonon coupling
since $v_{\rm ARPES}$ depends critically on the fit procedure.

\begin{figure}[t]
\includegraphics[width=7.25cm]{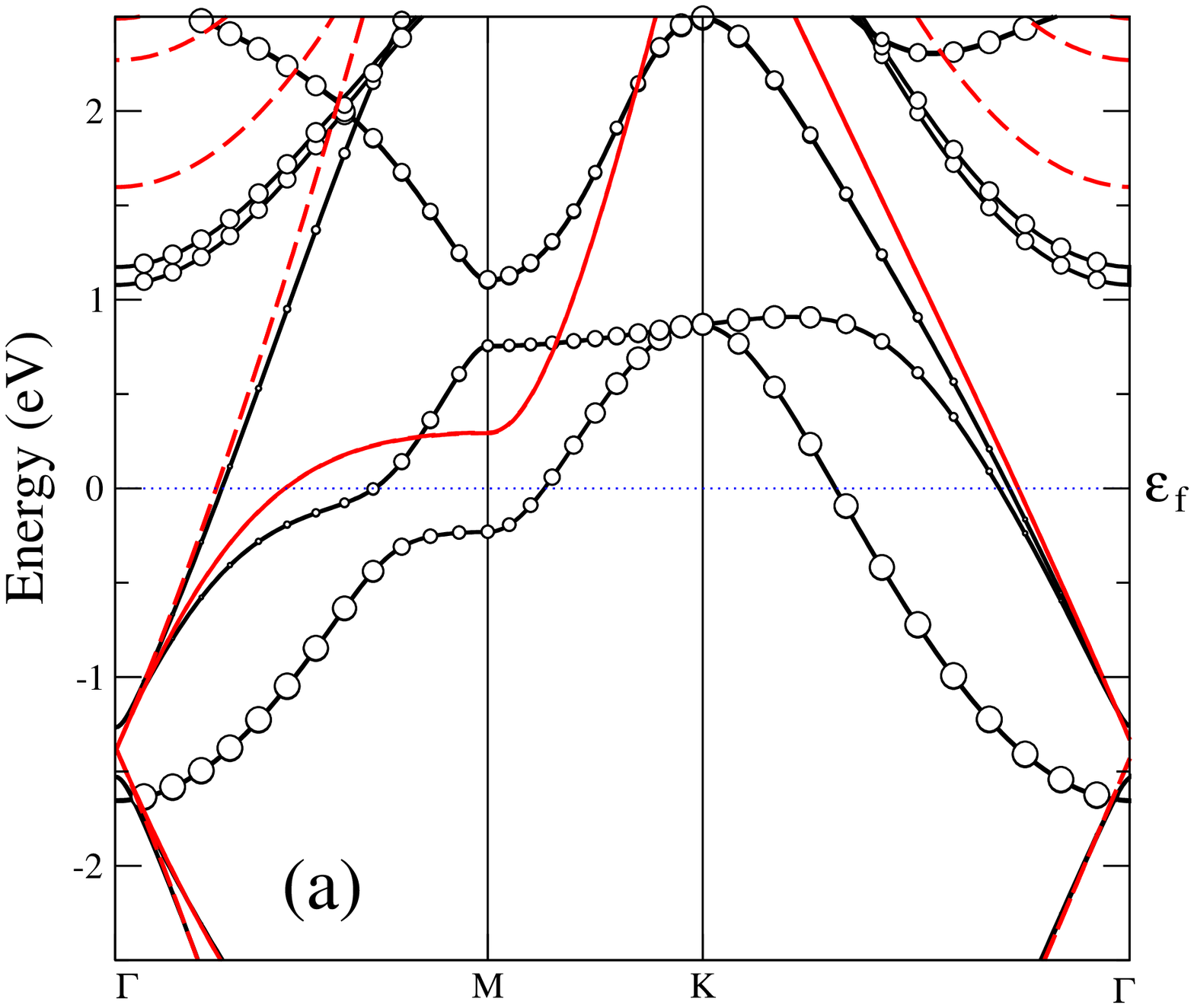}
\includegraphics[width=5.0cm]{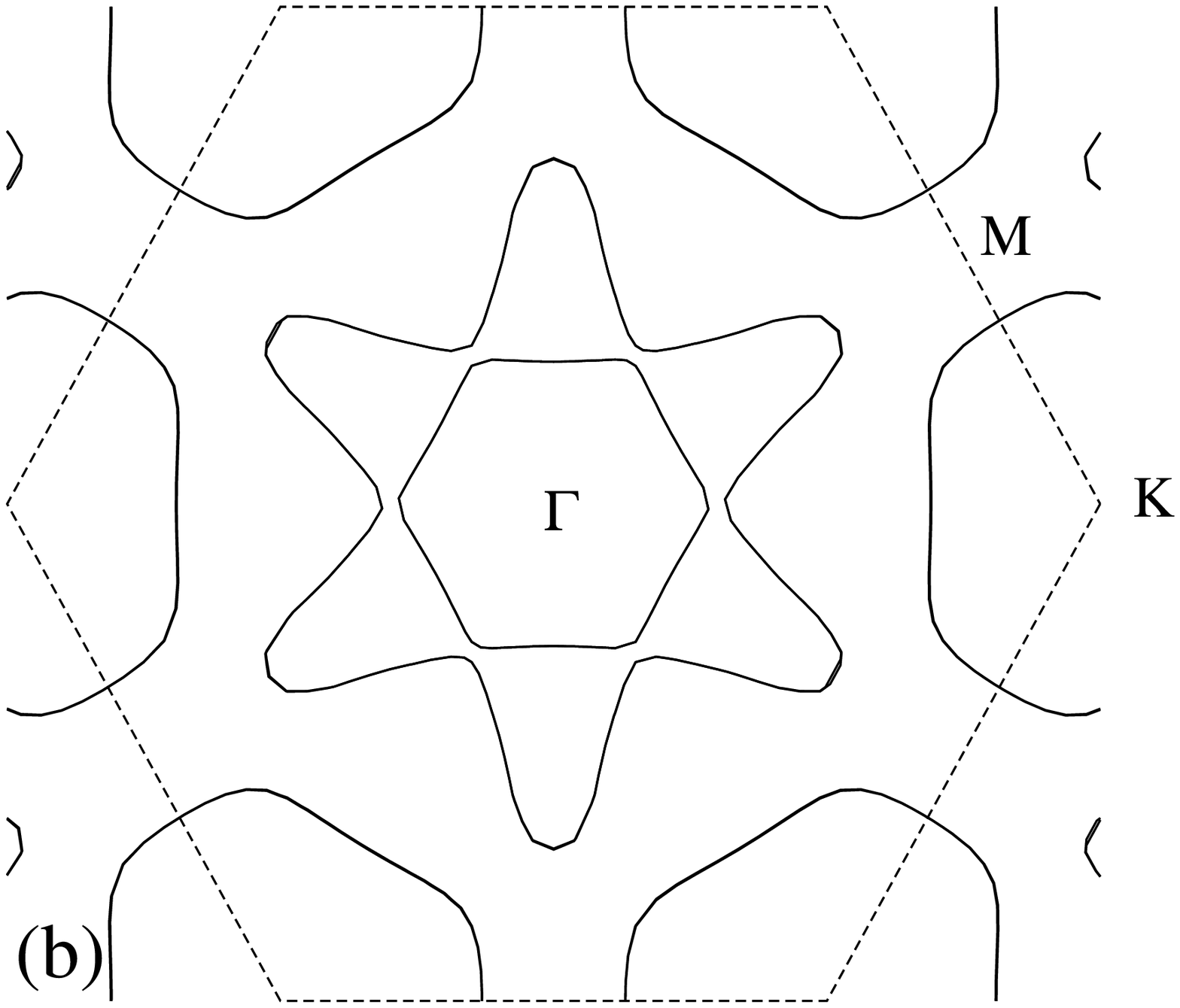}
\caption{(Color online) (a) Band structures of CaC$_6$ Monolayer
(continuous line) and graphene (red-dashed) using the same in-plane lattice parameter,
namely $a=2.47 $\AA\  (dashed). The size of the circles indicates the percentage
of Ca character (see ref. \cite{Calandra2005}) in a given band. 
(b) Fermi surface of 
CaC$_{6}$ monolayer.
The special points labels refer to the
CaC$_{6}$ hexagonal Brillouin zone.  }
\label{fig:Bands_mono_bulk}
\end{figure}

When large dopings are considered, as in the case of a CaC$_{6}$ monolayer
in ref. \cite{McChesney2007}, the commonly accepted interpretation 
of the band structure in term of rigid doping
of graphene $\pi^{*}$ states becomes questionable. 
For this reason we calculate the single particle band-structure of monolayer 
and bulk CaC$_{6}$ using density functional theory (DFT). 
We assume that the Ca deposition
leads to an ordered $\sqrt{3}\times\sqrt{3}$ structure with axes rotated of 30$^{o}$
respect to the standard C$_2$ crystal structure.
Simulating a single layer
requires very large cells to converge due to the finite electric dipole 
formed by the Ca donor and the graphene acceptor. The problem can be solved
using a double slab geometry with two specular layers of CaC$_6$  separated
by vacuum along the $z$ direction. This geometry has zero net electric 
dipole.
The interstitial space between Ca atoms on different layers is $11$ \AA\
while that between the two Graphene layers is $ 10$ \AA. 
We optimize the in-plane graphene 
lattice parameter ($a$) and the distance ($z$) between the Ca atoms and the nearby
graphene layer, obtaining $a=2.47$ \AA\ and  $z=2.315$ \AA.
Note that in bulk CaC$_{6}$ structural optimization leads to
$a=2.50$ \AA\ and $z=2.60$ \AA\ \cite{Calandra_GICs}.  
Electronic structure calculations are performed using
 the espresso code\cite{PWSCF} and
the generalized gradient approximation \cite{PBE}. We expand
the wavefunctions and the charge density using a 35 Ry
and a 600 Ry cutoffs, respectively. 
The electronic integration for the double slab geometry has been performed
using a $8\times 8\times 8$ k-points mesh and a Gaussian smearing of 0.05 Ryd.
For graphene we use a $30\times 30\times 2$ k-points mesh.

\begin{figure*}[t]
\includegraphics[width=6.75cm]{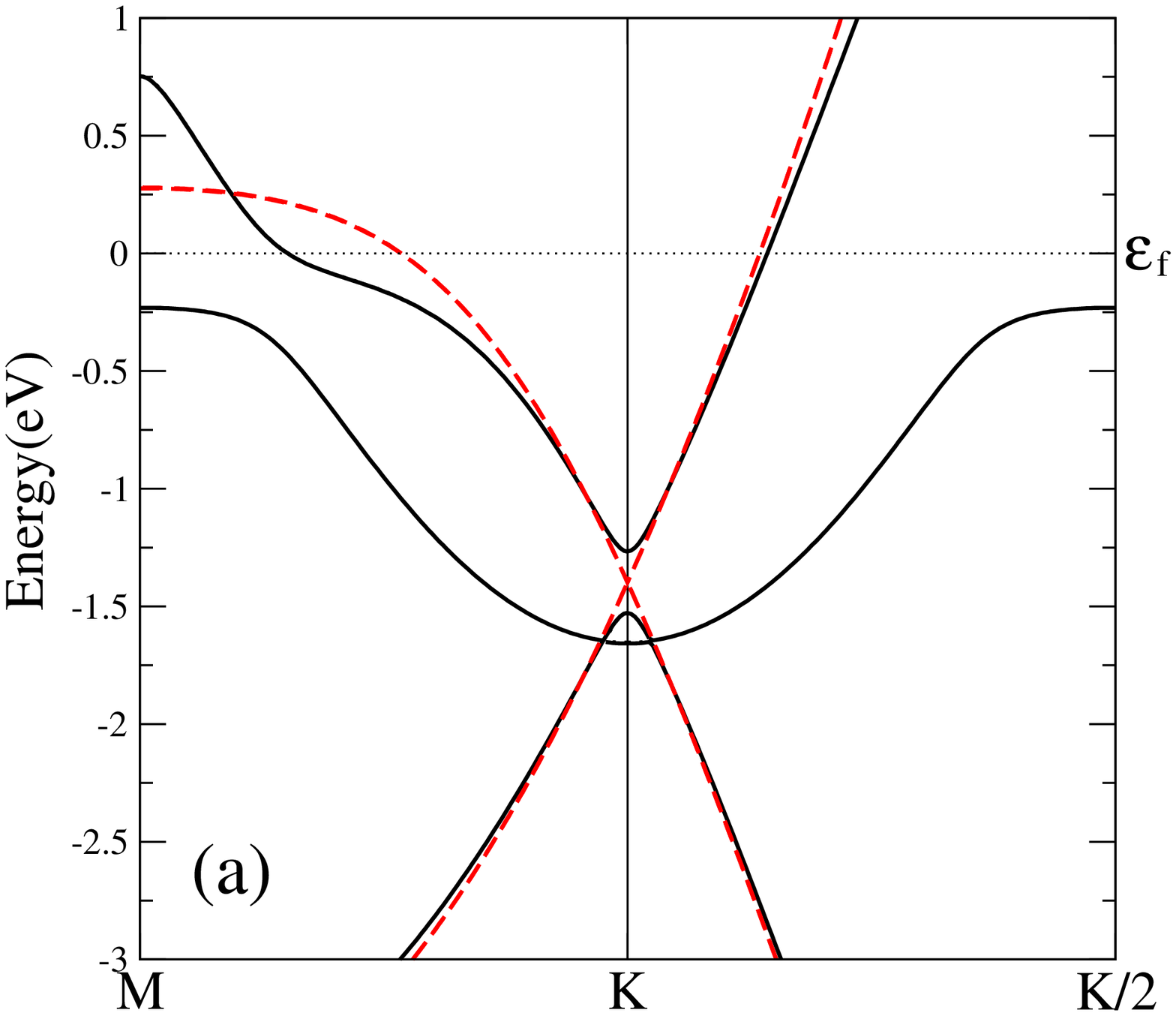}%
\includegraphics[width=5.75cm]{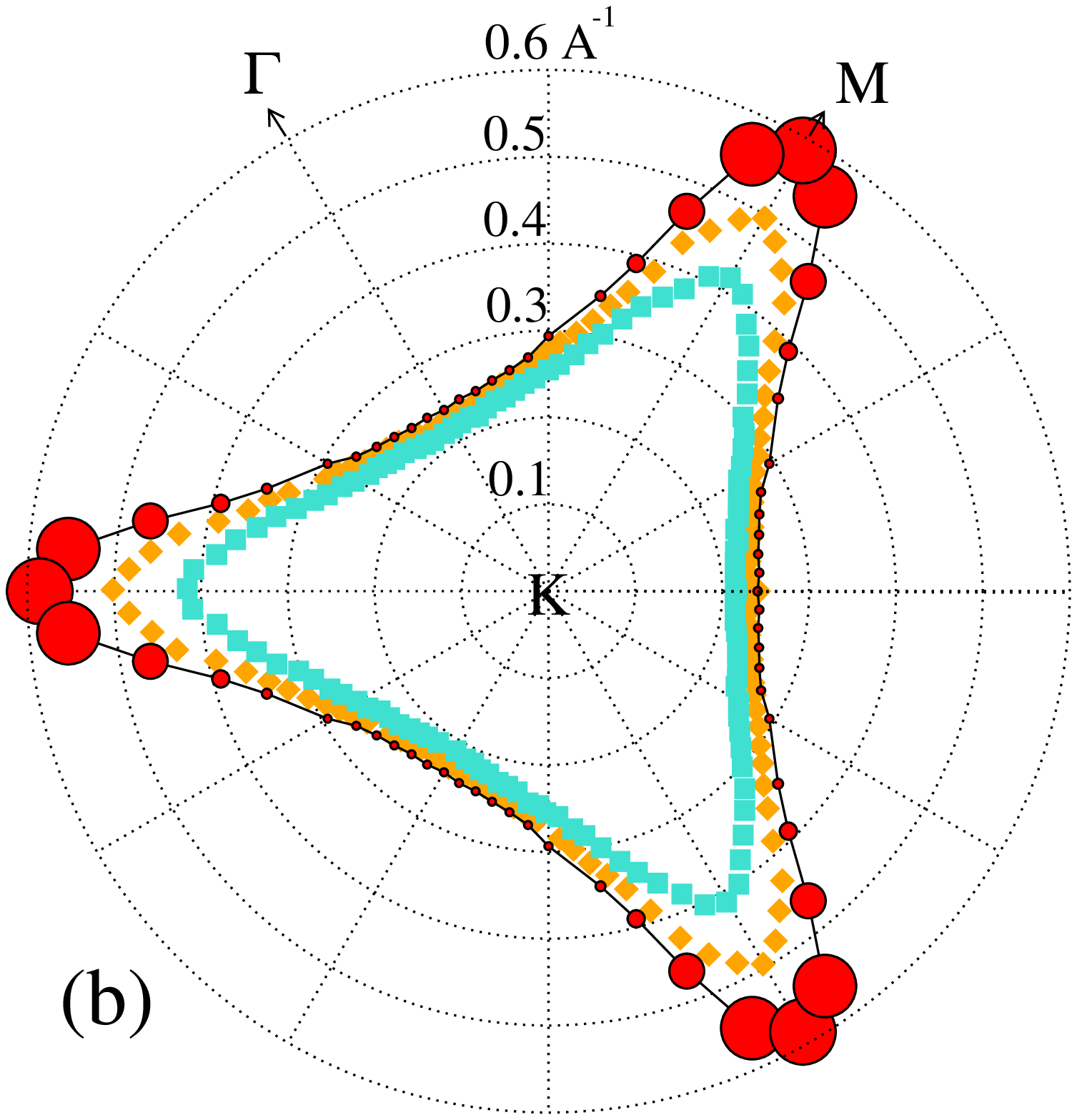}%
\includegraphics[width=6.0cm]{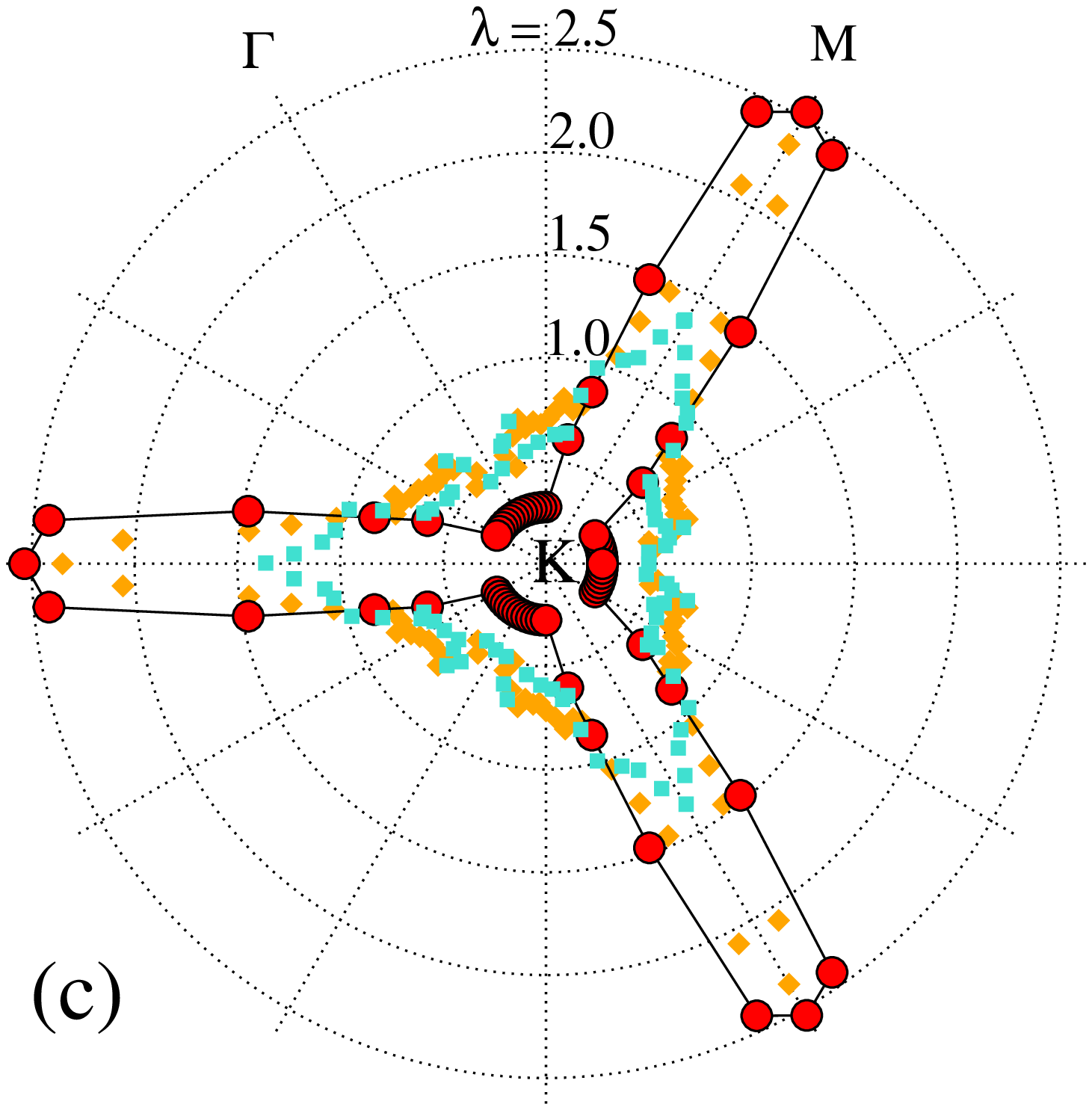}
\caption{(color online) (a) Band structure of CaC$_6$ Monolayer 
(continuous and dotted lines) and of Graphene (dashed lines) plotted 
in the Graphene Brillouin zone. (b) Fermi surface of CaC$_{6}$ as a function of doping. 
Square and diamonds are data from ref. \cite{McChesney2007} and refers to the dopings
labeled Ca (top) and Ca (under) K $2\times 2$ (top). 
Circles are theoretical calculations. 
The size of the circles is proportional to $\lambda_{\rm ARPES}$. (c)
Polar plot of $\lambda_{\rm ARPES}$. 
The high-symmetry points are labeled respect to the C$_{2}$ Brillouin
zone.}
\label{fig:Polar_Plots}
\end{figure*}

In fig. \ref{fig:Bands_mono_bulk} we show the band structure of CaC$_{6}$
monolayer compared to that of graphene using the same
in-plane lattice parameter, $a=2.47$ \AA. The energies of the 
Dirac point in the two structures have been aligned.
In the CaC$_{6}$ monolayer Brillouin zone the Dirac point is
refolded at $\Gamma$ due to the 
$\sqrt{3}\times\sqrt{3}$R$30^{o}$ surface deposition.
Because of this superstructure,
a small gap opens at the Dirac point in CaC$_6$ monolayer.
The Fermi level is crossed by two Carbon $\pi^{*}$ bands 
and a third band mainly Ca in character.
The steepest $\pi^{*}$ band (labeled $\pi^{*}_{1}$) is very similar to that of C$_{2}$,
while the other $\pi^{*}$ (labeled $\pi^{*}_{2}$) is 
substantially affected by the Ca recovering. 
The most striking feature is, however, the occurrence of an additional
band at $\epsilon_f$, having dominant Ca character. This band is the
so called ``intercalant'' band in GICs 
\cite{Csanyi,Calandra2005}, descending
to lower energies respect to its position in CaC$_{6}$ bulk.
The bottom of the intercalant band is at $-1.15$ eV respect to $\epsilon_f$
in bulk CaC$_{6}$ and is now at $-1.65$ eV in the monolayer. 
Correspondingly the graphene Dirac point is at $\approx -2.04$ eV in the 
bulk and at $\approx 1.39$ in the monolayer, in nice agreement with 
experiments\cite{McChesney2007}.
This effect is well known;
in alkaline-earths GICs, the intercalant band descends 
as the interlayer distance is reduced \cite{Csanyi,Calandra_GICs,Boeri}. 

Finally, in CaC$_{6}$ monolayer there is a marked avoided crossing
between the $\pi^{*}_{2}$ band and the Ca-band
along $\Gamma M$.
In this direction, approaching the M-point, there is a large
hybridization between the two bands, as it is evident
from the change in the Ca character of the band. 
Most important, a 
large deviation from linearity occurs in the $\pi^{*}_{2}$ band at
an energy of $\approx 0.2$ eV below the Fermi energy. 
This energy is very close to the 
E$_{2g}$ phonon frequency \cite{Lazzeri2006} of undoped graphene at the ${\bf \Gamma}$
point, namely  $\omega_{{\rm E}_{2g} {\bf \Gamma}}=0.195$ eV.

The Fermi surface of CaC$_{6}$ monolayer is shown in fig. \ref{fig:Bands_mono_bulk}
(b) in the CaC$_{6}$ monolayer Brillouin zone and in fig. \ref{fig:Polar_Plots}
(b) in the C$_2$ Brillouin zone. In fig. \ref{fig:Bands_mono_bulk} (b)
the Fermi surface is composed of a 6-points star centered at $\Gamma$
formed by the intersection of two triangular shapes. In addition
there are 6 hole pockets centered around ${\bf K}$ an ${\bf K'}$. 
The hexagon within the star is due to the $\pi^{*}_{1}$ band, 
while the outer perimeter of the star is composed by the 
$\pi^{*}_{2}$ band hybridized with Ca states.  
Along $\Gamma K$ the star has dominant C character, while
along  $\Gamma M$ it has 
strongly hybridized Ca and C character. 
The hole pockets are mostly Ca. 
In fig. \ref{fig:Polar_Plots} (b) one of the triangular shapes composing the
star is compared with the Fermi surface determined by ARPES \cite{McChesney2007}.
A very good agreement is found for the surface 
at the highest experimental doping.

In fig. \ref{fig:Polar_Plots},
the CaC$_{6}$ monolayer and C$_2$ bands are plotted in the C$_{2}$ 
Brillouin zone. 
The aforementioned  deviation from linearity in the $\pi^{*}_{2}$ bands,
now occurring along {\bf MK}, results in an apparent kink. 
However this kink is not due to the electron-phonon
interaction or to many body effects, but it is due to the
non-linearity of the single particle bands.
The change in slope is induced indirectly by the hybridization with
the intercalant band.

In order to compare with experimental data, 
we estimate the ``apparent'' electron-phonon coupling
$\lambda_{\rm ARPES}$ (Eq. \ref{eq:lambda_arpes}) generated by the
change in slope of the single particle band.
We obtain $ v_0$ and $ v_{\rm ARPES}$ in Eq. \ref{eq:lambda_eff} 
by linear fits to the $\pi^{*}_{2}$
band at several directions departing from the ${\bf K}$ 
point in the C$_{2}$ Brillouin zone. 
The fits to obtain $ v_{\rm ARPES}$ and $ v_0$ are
performed in the energy range $-0.2 {\rm eV} < \epsilon-\epsilon_f < 0 {\rm eV}$
and $-1.0 {\rm eV} < \epsilon-\epsilon_f < -0.2 {\rm eV}$, respectively.
A strong anisotropy in $\lambda_{\rm ARPES}$ is obtained as a result
of the single-particle band-structure. Values as large as 2.5 are found along
the ${\bf KM}$ direction. Along $\Gamma K$, where no kink is present in
the bare bands, a zero value is found. 

To compare with experiments the
contribution due to the electron-phonon coupling in the CaC$_{6}$ monolayer
should be added. In ARPES spectra in ref. \cite{McChesney2007}
the only visible contribution is that due to the in-plane graphene phonons
at $\approx 0.2$ eV. In bulk CaC$_{6}$ this contribution is 0.11 \cite{Calandra2005}.
As we have shown in ref. \cite{Calandra2007}, the fitting procedure used in
ref. \cite{McChesney2007} overestimates $\lambda$ of a factor 2.5. Thus
to the anisotropic bare band contribution we added an isotropic electron-phonon
contribution of 0.275.

The results are illustrated as polar plots in Fig. \ref{fig:Polar_Plots} (b)
and (c) and are compared to the experimental data of ref. \cite{McChesney2007}
in Fig. \ref{fig:Polar_Plots} (c).
We reproduce the strong anisotropy observed in ARPES experiment \cite{McChesney2007}
and the huge values of $\lambda_{\rm ARPES}$ along the ${\bf KM}$ direction.
In the ${\bf K\Gamma}$ direction, where the bare band contribution is 
zero, the value of $\lambda_{\rm ARPES}$ is given entirely by the electron-phonon
contribution. In this direction the values of $\lambda$ for bulk CaC$_{6}$
underestimates the experimental one. This could be due
a larger density of states at the Fermi level
in CaC$_{6}$ monolayer respect to bulk CaC$_{6}$. Indeed we found
that the monolayer density of states at the Fermi level is 1.7 time larger 
than that in the bulk.

In this work we have studied the electronic structure of 
foreign atoms deposed on graphene. We have shown that as the charge
transfer to graphene becomes significant, the band structure
cannot anymore be interpreted in terms of the pristine graphene
$\pi^{*}$ bands. When Ca is deposed on graphene at stoichiometry comparable
with CaC$_{6}$, a Ca band occurs at the Fermi level. 
This band is the surface analogue of the
intercalant band in superconducting graphite intercalated compounds
\cite{Csanyi,Calandra2005}, that is crucial to understand
the superconducting behavior of these compounds.
The Ca bands strongly hybridize with the C $\pi^{*}$ states and induces a
marked non-linearity in one of the $\pi^{*}$ bands at energies
comparable to $\epsilon_f-\omega_{{\rm E}_{2g}{\bf \Gamma}}$. This 
non-linearity explains the large and strongly anisotropic
values of $\lambda_{\rm ARPES}$ in CaC$_{6}$ monolayer. However,
as shown in this work, these large and anisotropic values do not provide any
information on the real electron-phonon coupling in the system.

We acknowledge illuminating discussions with Eli Rotenberg, I. Mazin,
J. McChesney and A. Bostwick.
Calculations
were performed at the IDRIS supercomputing center (project 071202).

\end{document}